\documentclass[preprint,12pt]{elsarticle}

\usepackage{amssymb}
\usepackage{amsmath}
\usepackage{longtable}
\usepackage{subcaption}
\usepackage{multirow}
\usepackage{footnote}
\usepackage{caption}
\usepackage{longtable}
\usepackage{amssymb}
\usepackage{pifont}
\makesavenoteenv{longtable}
\usepackage{longtable}
\usepackage{pifont}
\newcommand{\cmark}{\ding{51}} 
\newcommand{\xmark}{\ding{55}} 
\usepackage{multirow}
\usepackage{makecell} 
\usepackage[utf8]{inputenc}
\usepackage[T1]{fontenc}
\usepackage{array}  
\usepackage{lmodern} 
\usepackage{geometry}
\usepackage{graphicx} 
\usepackage{array} 
\usepackage{adjustbox} 

\journal{Economic Analysis and Policy}

\begin{document}

\begin{frontmatter}



\title{An Artificial Trend Index for Private Consumption Using Google Trends}




\author[inst1]{Juan Tenorio}
\affiliation[inst1]{organization={Universidad Peruana de Ciencias Aplicadas},
           addressline={2390 Prolongación Primavera}, 
          city={Santiago de Surco},
         postcode={15023}, 
        state={Lima},
       country={Peru, pcefjten@upc.edu.pe}}

\author[inst2]{Heidi Alpiste}
\affiliation[inst2]{organization={Universidad Peruana de Ciencias Aplicadas},
           addressline={2390 Prolongación Primavera}, 
          city={Santiago de Surco},
         postcode={15023}, 
        state={Lima},
       country={Peru, pcenhalp@upc.edu.pe}}

\author[inst3]{Jakelin Remón}
\affiliation[inst3]{organization={Pontificia Universidad Católica del Perú},
          addressline={1801 Universitaria}, 
        city={Cercado de Lima},
      postcode={15088}, 
     state={Lima},
    country={Peru, jakelin.remong@pucp.edu.pe}}

\author[inst4]{Arian Segil}
\affiliation[inst4]{organization={Universidad Nacional Mayor de San Marcos},
          addressline={375 Carlos Germán Amezaga}, 
        city={Cercado de Lima},
      postcode={15081}, 
     state={Lima},
    country={Peru, arian.segil@unmsm.edu.pe}}

\begin{abstract}
In recent years, the use of databases that analyze trends, sentiments or news to make economic projections or create indicators has gained significant popularity, particularly with the Google Trends platform. This article explores the potential of Google search data to develop a new index that improves economic forecasts, with a particular focus on one of the key components of economic activity: private consumption (64\% of GDP in Peru). By selecting and estimating categorized variables, machine learning techniques are applied, demonstrating that Google data can identify patterns to generate a leading indicator in real time and improve the accuracy of forecasts. Finally, the results show that Google's "Food" and "Tourism" categories significantly reduce projection errors, highlighting the importance of using this information in a segmented manner to improve macroeconomic forecasts.
\end{abstract}



\begin{keyword}
Neural Networks \sep  Private Consumption \sep Google Trends.
\end{keyword}

\end{frontmatter}


\section{Introduction}
\label{sec:introduction}

In the Peruvian economic context, private consumption has been a fundamental pillar of GDP, accounting for between 61\% and 66\% of the annual total since 2010 (see Figures A.6 and A.7). This figure highlights the importance of having precise and up-to-date information about household consumption behavior, as it allows for more accurate inferences about the dynamics of the country's economic activity. However, reality shows that consumption data, mostly available quarterly and monthly, face a considerable delay in publication, averaging 45 days after the close of the reference quarter. Due to this time delay, analysts and policymakers cannot respond promptly to changes in consumption trends, which can affect the effectiveness of economic decisions by not being made in a timely manner. 

To address these challenges, it is common to rely on leading indicators that provide information more immediately. These indicators include both quantitative data on observed spending decisions and qualitative information obtained from household surveys. However, even though many of these indicators are available monthly, they also face publication delays ranging from one to three months. In this sense, the situation calls for exploring new data sources that can offer more up-to-date estimates of private consumption.

In recent years, various studies have emphasized the potential of alternative data sources to monitor and provide early estimates of economic indicators, with a particular focus on real-time consumption. Among these sources are news from the media, electronic ATM payment data, and Google search queries, quantified through Google Trends. In this sense, recent studies, such as those by \cite{choi2012predicting} and \cite{bok2018macroeconomic}, have shown that these data can be useful for complementing and, in some cases, improving projections based on traditional indicators.

Therefore, this document focuses on the exploratory analysis of the advantages of using Google Trends as a tool to create a monthly indicator of private consumption. The choice of this source is based on the premise that online searches reflect consumers' intentions and concerns before they materialize in purchasing decisions. Thus, Google Trends offers the possibility of capturing early signals of changes in consumption behavior, giving analysts a certain advantage in anticipating and identifying trends.

In this way, the combination of Google variables and the use of advanced techniques such as artificial and recurrent neural networks help identify patterns that can be used to create an indicator and improve prediction accuracy. Additionally, the ability of neural networks to model complex and non-linear relationships between variables has proven particularly useful in detecting patterns in consumption, employment, tourism, among others. For instance, studies like those by \cite{hinton2012improving} and \cite{pascanu2013construct} have highlighted the effectiveness of these networks in uncovering underlying trends in large volumes of data, making them a valuable tool for developing economic indicators based on unconventional data.
 
The structure of this document is organized as follows: first, an exhaustive review of the literature is presented, exploring the relationship between sentiment variables derived from surveys and Google searches, as well as the use of neural networks for predicting consumption and other economic variables. Subsequently, the methodology employed is detailed, highlighting the use of neural networks for the creation of the Artificial Consumption Trend Indicator (ACTI). Finally, the results are presented, followed by a section dedicated to robustness tests, and concluding with the implications of the findings for their use in economic policy formulation.

\section{Literature review}
\label{sec:lite}

Over the last decade, the forecasting of consumption has become a central topic in economic research, given its crucial role in policymaking and business decision-making. Traditionally, consumer confidence indices have been tools for predicting household spending behavior. However, the growing availability of data from non-official (unstructured) sources and advances in statistical and econometric techniques have opened new possibilities for improving forecasts and creating new indicators. This section reviews the literature that has explored both traditional and innovative approaches to forecasting consumption, examining studies conducted in various geographic areas and economic contexts. Throughout this review, the strengths and limitations of different methods are identified, as well as new opportunities arising from the integration of new data sources and advanced techniques in this field.

First, numerous studies have examined the ability of consumer confidence indices to forecast consumption, generating extensive debate in this research area. In the United States, the MCSI (University of Michigan Consumer Sentiment Index) and the CCI (Conference Board Consumer Confidence Index), both survey-based, have frequently been used as key tools for capturing the dynamics of consumption. Furthermore, \cite{carroll1994does} demonstrated that, even with a time delay, the consumer confidence index can explain significant variations in household spending—an important finding in the field of behavioral economics.

Similarly, subsequent studies have extended this analysis of consumer confidence indices to other geographic contexts, revealing consistent patterns but also notable differences. For example, \cite{kwan2006usefulness} and \cite{li2016forecasting} applied these models in Canada, finding that the Conference Board confidence indices are not only effective for predicting total consumption spending, but also its subcategories, such as durable goods, non-durable goods, and services. Additionally, in Nigeria, \cite{olowofeso2012consumer} identified that both the consumer confidence index and the retail trade index are significant variables in capturing the trend of private consumption.

However, it is important to note that other studies have highlighted certain limitations of these indicators, fueling debate about their true effectiveness. Specifically, research indicates that their predictive capacity is significant only in specific periods and under certain conditions. Authors such as \cite{desroches2002usefulness}  and \cite{dees2013consumer}  conclude that, in the Eurozone and the United States, survey-based confidence indices are only relevant during periods of geopolitical tensions, financial crises, or high volatility. Such events alter consumer perceptions and significantly influence consumption, underscoring the need to consider the macroeconomic context when interpreting these indicators. Along the same lines, \cite{carrera2019consumption} found that, in the case of Peru, these indices tend to have limited influence on consumption behavior during stable periods, but their relevance increases considerably in times of uncertainty. Nonetheless, \cite{croushore2005consumer} and \cite{kwan2006usefulness} argue that, both in the Eurozone and the United States, the out-of-sample predictive capacity of these indices is restricted, especially when real-time data are used, raising questions about their practical utility beyond academic research.

In response to these limitations and capitalizing on technological advances, the creation of indicators and consumption forecasting has evolved with the inclusion of unstructured data, such as those provided by Google Trends, representing a significant innovation in economic analysis methodologies. \cite{vosen2011forecasting} constructed a new indicator of private consumption in the United States based on Google Trends searches, comparing its predictive power with the traditional MCSI and CCI indicators. Their results show that models augmented with Google Trends variables significantly outperform models based on survey-derived variables, especially in out-of-sample predictions. This approach suggests that online search data can offer a more up-to-date and accurate perspective of real-time consumption trends.

Similarly, \cite{woo2019forecasting} incorporated Google Trends data related to consumption and news to improve private consumption forecasts in the United States, broken down into durable goods, non-durable goods and services. Their findings indicate that the resulting models provide significant additional information, outperforming the predictive capacity of traditional consumer confidence indicators. In addition, data from news sources provide insight into changes in durable goods consumption, suggesting that this new information can complement and, in some cases, surpass traditional economic data sources.

In Spain, \cite{gil2018nowcasting} carried out a comparative study using various indicators to forecast short- and medium-term private consumption. These included both quantitative and qualitative traditional indicators, credit and debit card data, uncertainty indicators, and indicators derived from Google Trends. Their findings showed that Google-based indicators provide significant added value when combined with traditional and uncertainty indicators, as they capture purchase intentions before these intentions translate into actual transactions. They provide forward-looking indicators of potential shifts in consumer behavior, which can help analysts anticipate market trends. 

Likewise, in China \cite{song2023predicting} analyzed behavioral data from Baidu, the leading search platform in the country, to enhance consumption forecasts at both the sectoral and aggregate levels. Their findings indicate that incorporating Baidu-derived information substantially improves the accuracy of forecasts and reduces projection errors. This suggests that data from local search platforms can be a valuable source of economic insights in emerging markets, where traditional data sources may be less reliable or underdeveloped.

In the Peruvian context, recent studies have also begun to explore unstructured data for enhancing economic estimates. For instance,\cite{chang2013google} employed Google Trends to construct an index predicting aggregate employment in Peru. They found that the index forecasts both real-time analysis and one-period-ahead projections, although it does not anticipate changes beyond that horizon. This research illustrates how modern techniques can supplement traditional economic data. Similarly, in Chile,\cite{swallow2010nowcasting} developed an index of interest in automobile purchases based on Google search data and integrated it into macroeconomic forecasting models, obtaining more accurate projections. Finally, \cite{blanco2014herramientas} investigated whether combining Google Trends with other indicators in Argentina could improve the predictive capacity of various economic variables.

In terms of methodology, researchers have increasingly explored advanced data analysis techniques, such as neural networks, to develop indicators and estimate macroeconomic variables. For example, \cite{aydin2015comparison} compared the forecast performance of vector autoregressive (VAR) models with repeated neural networks (RNN), using variables such as gold prices, the Istanbul Stock Exchange index, and exchange rates. They found that RNNs generate more accurate estimates, suggesting that these models can be a powerful tool for economic forecasting, at times surpassing traditional econometric approaches. In the same way, \cite{zahedi2015application} investigated the potential to forecast stock prices on the Tehran Stock Exchange, employing Artificial Neural Networks (ANN) and Principal Component Analysis (PCA), and observed that both techniques effectively predict the variables under consideration. These findings reinforce the notion that machine learning–based statistical methods can complement and enhance traditional approaches in economic analysis.

In the case of inflation,\cite{barkan2023forecasting} employed an innovative RNN model called Hierarchical Recurrent Neural Networks (HRNN). They found that this model outperforms other traditional models based on evaluation metrics, highlighting the potential of these methodologies in macroeconomics. In addition, \cite{monge2024chinese} used a combination of techniques, such as text mining and factor analysis, together with data from Google Trends to create an indicator called RT-LEI, which delivers even more accurate GDP forecasts than the OECD’s Composite Leading Indicators (CLI). Finally, \cite{tenorio2024gdp} combined leading indicators with Google Trends search data to project the monthly GDP of Peru, observing a significant reduction in forecast error when these variables were included. This suggests that integrating diverse data sources and analytical methods can lead to substantial improvements in economic predictions.

In summary, the evolution of techniques for creating indicators and estimating consumption reflects a shift from traditional models based on consumer confidence to more sophisticated approaches that integrate unstructured data and advanced analytical methods. Furthermore, the studies reviewed emphasize how the incorporation of modern data sources, such as online search queries, along with the application of machine learning not only improved the accuracy of the forecast, but also provided more timely insights into consumption trends. However, these innovations are not without challenges, particularly in terms of the interpretation and application of these models in diverse macroeconomic contexts.


\begin{longtable}{p{4cm} p{4cm} p{4cm}}
\caption{Literature review on the creation of indicators using Google Trends}
\label{tab:literature-review}\\
\hline 
\multicolumn{1}{c}{\textbf{Author}} & \multicolumn{1}{c}{\textbf{Index}} & \multicolumn{1}{c}{\textbf{Country}} \\ \hline
\endfirsthead

\multicolumn{3}{c}{{\tablename} \thetable{} -- Continued} \\
\hline 
\multicolumn{1}{c}{\textbf{Author}} & \multicolumn{1}{c}{\textbf{Index}} & \multicolumn{1}{c}{\textbf{Country}} \\ \hline
\endhead

\hline 
\multicolumn{3}{r}{{Continued on next page}} \\
\endfoot

\hline \hline
\multicolumn{3}{p{12cm}}{\small Source: Own elaboration.} \\
\endlastfoot

Larrahondo, Díaz, and Guerrero (2024) & Monthly arrivals of international tourists & Peru, Ecuador, Colombia, Bolivia  \\ 
\hline
Lolic, Matosec, and Soric (2024) & Monthly retail sales & United States \\ 
\hline
Woloszko (2020) & Weekly economic activity & Organization for Economic Cooperation and Development (OECD) \\ 
\hline
Santana (2020) & Monthly economic activity & Dominican Republic \\ 
\hline
Blanco (2014) & Monthly auto sales and consumption & Argentina \\ 
\hline
Chang and Del Río (2013) & Monthly employment in Lima for firms with 100 or more workers & Peru \\ 
\hline
Carriere-Swallow and Labbé (2010) & Monthly auto sales & Chile \\ 

\end{longtable}

\section{Methodology}
\label{sec:meth}
This section describes the treatment of Google Trends data used to gather the information for the creation of the ACTI and the prediction of private consumption, applying various methodologies. Monthly data from January 2007 to October 2024 is employed.  The main techniques for developing the ACTI are \textit{Artificial Neural Networks} (ANN) and \textit{Recurrent Neural Networks} (RNN). Benchmark models are also used to validate its implementation. Finally, the calibration of the networks is detailed through cross-validation.

\subsection{Google Trends data}
Google Trends is an online tool developed by Google to measure the popularity of specific search terms over time. It provides an index ranging from 0 to 100, representing the proportion of searches for a particular term relative to the total search volume within a specific region and time frame. Additionally, Google Trends offers data with daily, weekly, monthly, or even hourly frequency, allowing for detailed analysis tailored to the user's needs. 

This index does not reflect the absolute number of searches but rather a relative measure that facilitates comparison between terms and the identification of trends. The tool distinguishes between two types of data: (i) real-time data, representing a random sample of searches from the past seven days, and (ii) historical data, an independent random sample covering searches since 2004. Consequently, Google Trends is widely used in economic research to analyze behavior and consumption patterns derived from online searches, enabling a more precise understanding of current and emerging trends \cite{camusso2021google, gil2018nowcasting}. 

The trend is generated through the following steps: (i) dividing the number of searches for the term by the total number of searches within the selected region and/or time period to prevent cities with higher search volumes from dominating the results; and (ii) indexing the data on a scale from 0 to 100 by dividing each value by the highest result of the previous weight and multiplying it by 100. This process provides a time series of the search volume index for terms entered into Google. Given its utility and the high-frequency data offered by Google Trends, this research introduces the Artificial Consumption Trend Indicator (ACTI), a private consumption index based on Google search data 

To analyze search trends, the methodology involved extracting and transforming data from Google Trends. The process began with selecting a set of approximately 130 key terms (see Table 5), identified through a bibliographic review and deemed relevant for the study. Historical monthly data for these terms, from 2007 to 2024, was extracted using \textit{Pytrends}, Python interface for the Google Trends API.  The initial data was obtained in levels, representing the raw search volume for each term during each time period. To improve interpretation and analysis of temporal fluctuations, logarithmic transformations were applied. This logarithmic transformation normalizes time series, reducing noise in percentage variations over time regardless of the absolute search levels.

Subsequently, logarithmic variations of the series were calculated to capture relative changes more robustly against seasonal fluctuations or anomalous search spikes. This approach is particularly useful in economic analysis, as it facilitates the identification of behavioral patterns over time and comparison across terms with varying search volumes. This methodology ensures that the results accurately reflect underlying trends and are not distorted by inherent volatility in Google Trends data. Finally, for the specific creation of the ACTI, 26 terms relevant to private consumption and 23 terms related to commerce and services (see Table 6) were selected. Additionally, the choice of words aligns with the purposes of private consumption according to INEI (see Table 7). Statistical techniques (see Figures 8 and 9) and analytical methods were employed to validate the appropriate selection of search terms, capturing the best matches. This strategic selection ensures that the analysis aligns with major consumption trends, providing a reliable indicator for capturing patterns.

\subsection{Principal Component Analysis (PCA)}

Principal Component Analysis (PCA) is a dimensionality reduction technique widely used in data analysis, especially when dealing with a large number of correlated variables. Its main objective is to transform a set of observed variables into a new set of uncorrelated variables, known as principal components, which explain most of the variability in the original data. 

Unlike Artificial Neural Networks (ANNs), which aim to capture complex nonlinear relationships between variables, PCA is a linear method that focuses on identifying the directions in which the data vary the most. These directions correspond to the eigenvectors of the data's covariance matrix, while the associated eigenvalues indicate the amount of variance explained by each component.

The transformation of data through PCA is performed as follows:

\begin{equation}
    Z = XW
\end{equation}

where \( X \) is the matrix of original data, \( W \) is the matrix of eigenvectors, and \( Z \) is the resulting matrix of principal components. The principal components are ordered such that the first component explains the largest portion of the variance, followed by the second, and so on.

PCA has the advantage of simplifying the model by reducing the number of variables required to capture most of the information contained in the original data. This methodology is particularly useful in scenarios where the number of predictors is high relative to the number of observations, as it helps to prevent overfitting and multicollinearity issues. 

In contrast to neural networks, which can handle nonlinear and complex relationships, PCA is better suited when linear relationships among variables are sufficient to capture the underlying structure of the data. In this study, PCA is used to select an optimal subset of predictors, which are then employed in private consumption projection models.

\subsection{Dynamic Factor Models (DFM)}

Dynamic Factor Models (DFM) are a statistical tool used to model multivariate time series. These models are particularly useful when working with large sets of macroeconomic data, as they allow the joint dynamics of multiple variables to be captured through a small number of unobserved factors.

DFM assumes that the temporal evolution of a wide set of time series can largely be explained by a few common factors, which are dynamic over time. Mathematically, a DFM is represented as:

\begin{equation}
    X_t = \Lambda F_t + \epsilon_t
\end{equation}

where \( X_t \) is the vector of observations at time \( t \), \( F_t \) are the latent (unobserved), \( \Lambda \) is the factor loading matrix that relates the factors to the observations, and \( \epsilon_t \) is the specific error term for each time series. 

One of the main advantages of DFMs is their ability to synthesize the information contained in a large number of time series, capturing their co-movements through a few factors. This is particularly useful in the context of economic forecasting, where aggregating information can improve prediction accuracy. 

Unlike methodologies such as Neural Networks, which explore nonlinear and complex relationships, DFMs are linear and designed to capture the common underlying structure over time in a set of time series. In this study, DFMs are used as a benchmark model for creating the indicator and subsequently projecting private consumption, comparing their performance with other, more complex methodologies.

\subsection{Stepwise Least Square}
Stepwise regression is a method that sequentially adjusts a model by iteratively adding or removing variables based on different statistical criteria, aiming to minimize the mean squared error. This model successfully combines simplicity and determination to enhance the model's forecasting capability. The variable selection process can be performed using two approaches: forward selection and backward selection. Additionally, there is a combination of both approaches known as bidirectional stepwise regression. This research aims to find the best selection within the universe of 23 main indicators to predict private consumption.

\subsection{Artificial Neural Networks}

Artificial Neural Networks (ANN) are computational models inspired by the structure of the human brain, designed to recognize complex patterns and make accurate predictions. An ANN consists of multiple layers of interconnected neurons, where each neuron applies an activation function to a linear combination of the input received. These connections are weighted, and the weights are adjusted during the model's training process. 

In particular, neural networks are known for their ability to capture nonlinear relationships between input and output variables, making them especially useful in problems where traditional linear methods fail. Hidden layers allow the model to learn complex and hierarchical features from the data. 

The optimization process of ANNs is carried out by minimizing a loss function, commonly the mean squared error (MSE) in regression problems, using algorithms such as backpropagation along with an optimizer like the Adam method. The loss function for an artificial neural network is expressed as:

\begin{equation}
 \min_{\theta} \left( \frac{1}{n} \sum_{i=1}^{n} (y_i - f(\mathbf{x}_i; \theta))^2 \right)
\end{equation}

where \( \mathbf{x}_i \) represents the inputs, \( y_i \) the observed outputs, and \( \theta \) denotes the set of parameters (weights and biases) of the network.

ANN training involves iterative updating of the parameters \( \theta \) by error backpropagation from the output layer to the previous layers, thus adjusting the weights to reduce the prediction error. This approach enables the neural network to learn the underlying patterns in the data, enhancing its predictive capacity.

The flexibility and power of ANN make them a valuable tool for time series modeling and other economic applications, where the relationship between variables can be highly nonlinear and complex.

\begin{figure}[!ht]
\centering
\begin{minipage}[t]{1\linewidth}
\centering
\caption{{Processing diagram of a artificial neural network}}
\includegraphics[width=0.65\linewidth]{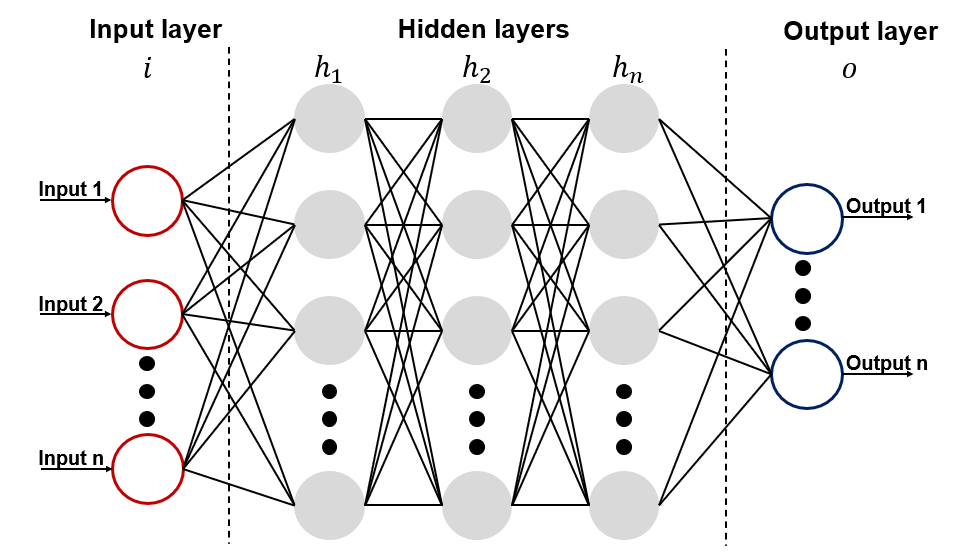}
\end{minipage}\\
\raggedright
\footnotesize{Source: \cite{bre2018prediction}}
\end{figure}

\subsection{Recurrent Neural Networks}

Recurrent Neural Networks (RNN) are an extension of Artificial Neural Networks (ANN) specifically designed to process sequences of data, making them particularly useful for tasks involving time series or sequential data. Unlike traditional ANN, where neurons in each layer are only connected to the next layer, RNN feature recurrent connections that allow a neuron's output to be fed back as input to the same neuron in subsequent time steps.

This structure enables RNN to "remember" previous information in the data sequence, which is crucial for capturing long-term temporal dependencies. The ability to retain information over time is one of the main differences from traditional ANN, which lack internal memory and treat each input instance independently.

The equation that defines the hidden state \( h_t \) in an RNN is as follows:

\begin{equation}
 h_t = f(W_h \cdot h_{t-1} + W_x \cdot x_t + b)
\end{equation}

where \( h_t \) is the hidden state at time \( t \), \( h_{t-1} \) is the hidden state at time \( t-1 \), \( x_t \) is the input at time \( t \), \( W_h \) and \( W_x \) are the weight matrices, and \( b \) is the bias. The function \( f \) is typically a nonlinear activation function such as tanh or ReLU.

A variant of RNN that has gained popularity is Long Short-Term Memory (LSTM), which introduces special mechanisms (such as input, forget, and output gates) to mitigate the vanishing gradient problem. This allows the network to learn long-term temporal dependencies more effectively.

\begin{figure}[!h]
\centering
\begin{minipage}[t]{1\linewidth}
\centering
\caption{{Processing diagram of a recurrent neural network (RNN)}}
 \includegraphics[width=0.65\linewidth]{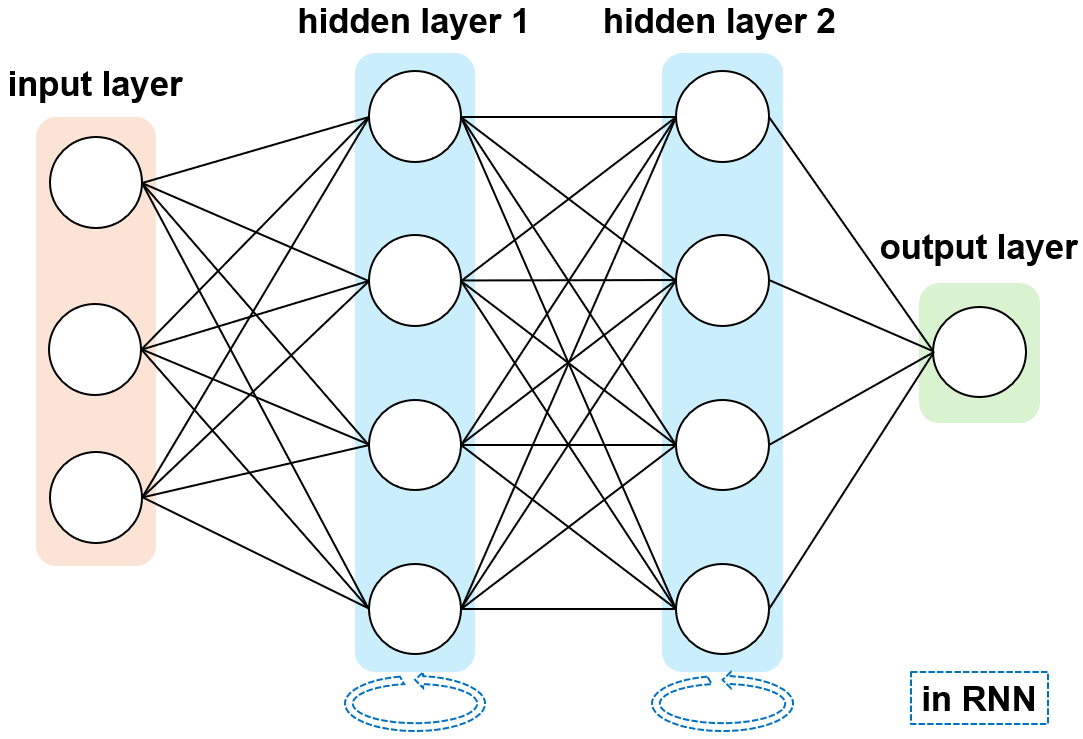}
\end{minipage}\\
\raggedright
\footnotesize{Source: \cite{ma2019classification}}
\end{figure}

In summary, while ANN are suitable for problems where the relationships between variables are primarily static, RNN and particularly LSTMs are more effective when capturing the temporal dynamics of data, such as in time series forecasting.

\subsection{Projection Evaluation Strategy}

The accuracy of each model's projection will be assessed using both the mean squared error (MSE) and the root mean squared error (RMSE). The MSE is calculated using the following equation:

\begin{equation} MSE = \frac{1}{T} \sum_{t=1}^{T}(y_t - \hat{y}_t)^2 \end{equation}

where $y_t$ represents the observed value of the monthly private consumption growth, $\hat{y}_t$ is the projected value, and $T$ is the total number of projections performed.

Meanwhile, the RMSE is obtained by taking the square root of the MSE:

\begin{equation} RMSE = \sqrt{\frac{1}{T} \sum_{t=1}^{T}(y_t - \hat{y}_t)^2} \end{equation}

Both statistics are used to measure the accuracy of the models, but they present key differences: while the MSE provides a measure of the average magnitude of the errors by squaring them, the RMSE returns the errors in the same units as the projected variable, making interpretation easier.

Once this initial evaluation of the prediction fit has been conducted, the method proposed by \cite{diebold1995paring} (DM) and the test by \cite{giacomini2006tests} (GW) will be employed to compare the differences in losses between two predictive models. The DM statistic is defined as:

\[
DM = \frac{\bar{d}}{\sqrt{\frac{\hat{\sigma}_d^2}{T}}}
\]

where \( \bar{d} \) is the mean of the differences in losses, \( \hat{\sigma}_d^2 \) is the estimated variance of these differences, and \( T \) is the number of observations. Additionally, the GW test was used to evaluate the conditional predictive accuracy of the models under different out-of-sample scenarios. Its statistic is defined as:

\[
GW = \frac{1}{\sqrt{n}} \sum_{t=1}^{n} \hat{d}_t
\]

where \( \hat{d}_t \) represents the difference in losses between the two models at time \( t \), and \( n \) is the size of the out-of-sample dataset.

\section{Results}
\label{sec:results}

This section presents the results obtained for each indicator, along with an analysis of the term selection based on explained or conditional variance, considering the prediction errors from subsamples. After generating an indicator for each methodology, cross-validation is performed to optimize the parameters used by the model. From this analysis, it is concluded that neural network models, both artificial and recurrent, effectively capture a consistent pattern that is robustly manifested across the various private consumption data sets. 

\begin{figure}[!h]
\centering
\caption{Private Consumption vs. Indicators by Methodology}
\noindent 
\begin{minipage}[c]{.5\textwidth}
  \centering
  \includegraphics[width=1\linewidth]{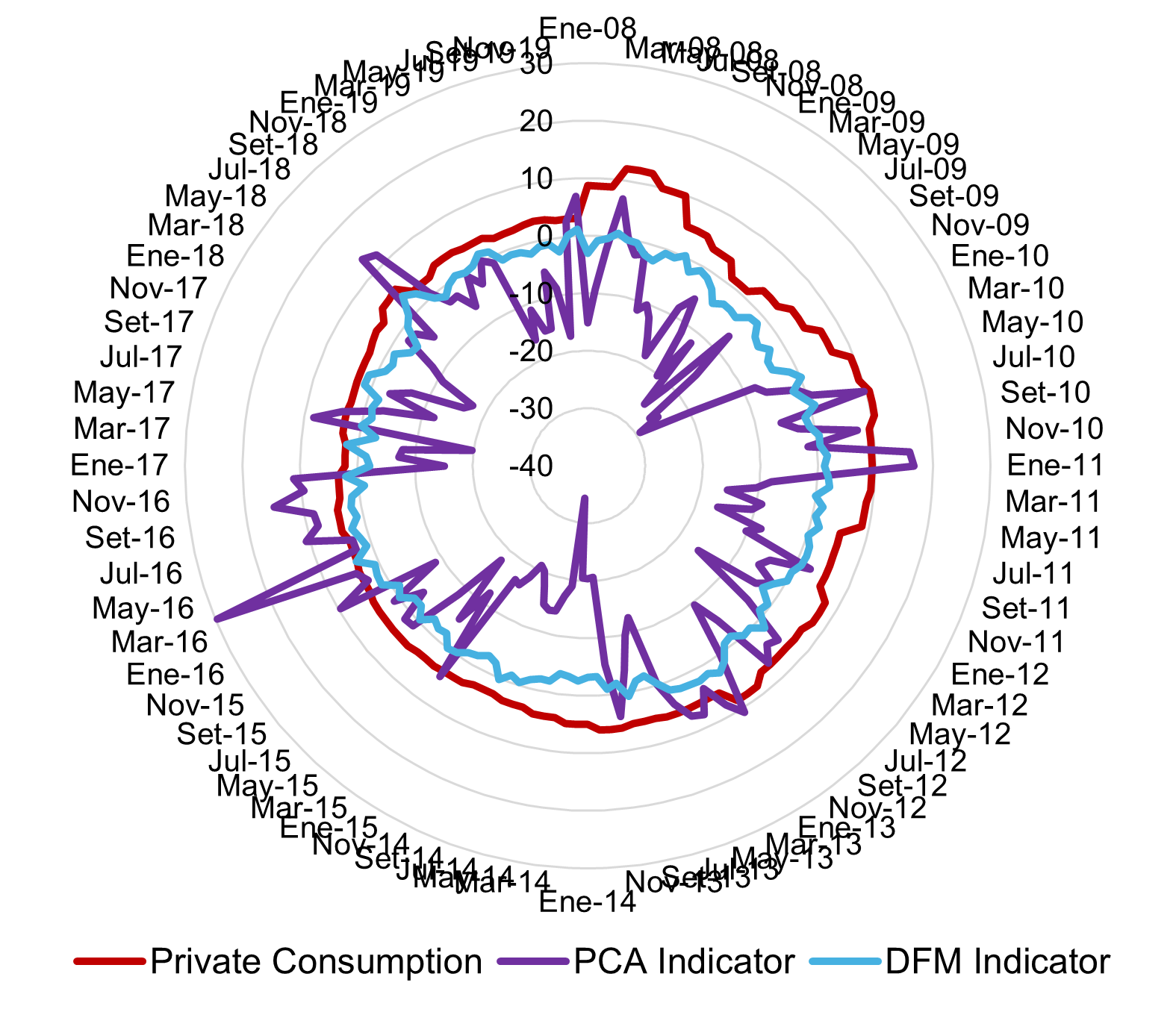}
  \subcaption{Private Consumption vs. Benchmark}\label{fig:1a}
\end{minipage}%
\begin{minipage}[c]{.5\textwidth}
  \centering
  \includegraphics[width=1\linewidth]{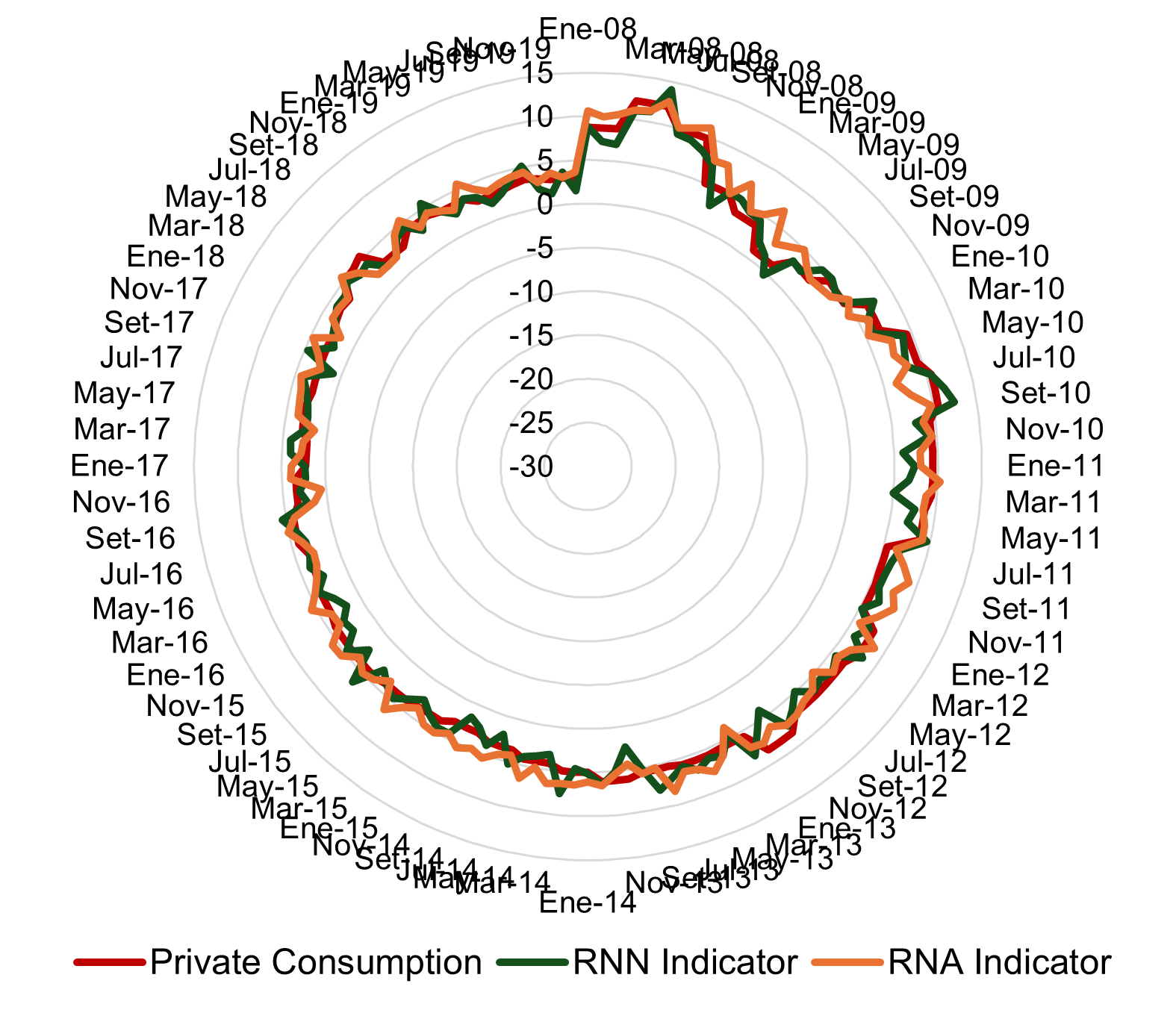}
  \subcaption{Private Consumption vs. Network Indicators}\label{fig:2a}
\end{minipage}
\raggedright
\footnotesize{Source: Own elaboration}
\end{figure}

In figure (3a), the indicators obtained through PCA, DFM, and private consumption are compared. It is observed that the PCA indicator exhibits significant fluctuations, while the DFM indicator shows a smoother behavior, suggesting that it better captures certain underlying trends, although it does not adequately fit the behavior of private consumption. Similarly, in figure (3b), the indicators from RNN and ANN are compared with private consumption. Here, both the RNN and ANN indicators closely follow the pattern and seasonal trend, with slight variations, indicating that these models more accurately capture the dynamics of the data. Therefore, it is evident that neural network models, such as RNN and ANN, are better equipped to adapt to complex and nonlinear fluctuations, excelling in capturing seasonal patterns and adjusting to unexpected changes in the data. However, it is crucial to emphasize that the exclusive use of unstructured data from Google Trends searches is essential for obtaining reliable results in the projection of private consumption.

\begin{figure}[!h]
\centering
\caption{{Volatility Analysis by Indicator}}
\noindent 
\begin{minipage}[c]{.5\textwidth}
  \centering
  \includegraphics[width=.95\linewidth]{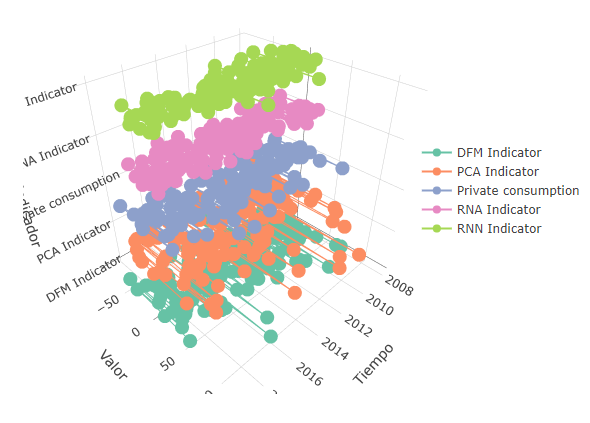}
  \subcaption{Conditional Variance $CVar(\epsilon)$}\label{fig:1a}
\end{minipage}%
\begin{minipage}[c]{.5\textwidth}
  \centering
  \includegraphics[width=.75\linewidth]{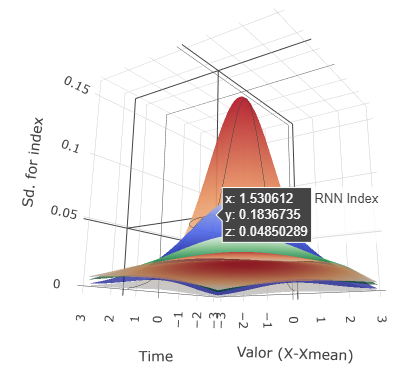}
  \subcaption{Varianza $Var(\epsilon)$}\label{fig:2a}
\end{minipage}
\raggedright
\footnotesize{Source: Own elaboration}
\end{figure}

\subsection{Hyperparameter Calibration}

Once the effectiveness of neural networks in identifying the best pattern for private consumption has been established, cross-validation (CV) will be employed in the neural networks and benchmarks to find the optimal parameter set for each method. To achieve this, the sample will be divided into three parts: i) Training (Jan 2008 to Aug 2014) with 5 folds, ii) Validation (Sep 2009 to May 2022), and iii) Testing (Jun 2022 to Oct 2024).

\begin{table}[h]
\centering
\caption{Cross-Validation Strategy for Parameter Optimization}
\begin{adjustbox}{center, max width=\textwidth}
\begin{tabular}{ccccccc}
\hline\hline
\multicolumn{5}{c}{\textbf{Training dataset}} & \multicolumn{1}{c}{\textbf{Validation set}} & \multicolumn{1}{c}{\textbf{Testing set}} \\ 
\multicolumn{5}{c}{2008m1-2014m08} & \multicolumn{1}{c}{2014m09-2022m5} & \multicolumn{1}{c}{2022m06-2024m10} \\ 
\hline\hline
\multicolumn{5}{c}{\textbf{$\longleftrightarrow$}} & \multicolumn{1}{c}{\textbf{$\leftrightarrow$}} & \multicolumn{1}{c}{\textbf{$\leftrightarrow$}}\\
Fold 1 & Fold 2 & Fold 3 & Fold 4 & Fold 5 &  \\ 
\hline\hline
\multicolumn{7}{p{\linewidth}}{\footnotesize\centering Source: Own elaboration} 
\end{tabular}
\end{adjustbox}
\end{table}

This approach will allow for the optimization of each model's hyperparameters, ensuring that any methodology not only fits the training data well but also generalizes effectively to new data. The goal is to prevent overfitting, thereby enhancing the stability and accuracy of each model across different prediction scenarios.

\setlength{\LTleft}{0.04\textwidth}
\setlength{\LTright}{0.04\textwidth}

\begin{longtable}{%
  >{\centering\arraybackslash}p{0.125\textwidth}%
  >{\centering\arraybackslash}p{0.24\textwidth}%
  >{\centering\arraybackslash}p{0.24\textwidth}%
  >{\centering\arraybackslash}p{0.22\textwidth}}
\caption{Priors and Range of Hyperparameters}
\label{tab:hyperparams}\\
\hline
\multicolumn{1}{c}{\textbf{\makecell{Modelo}}} &
\multicolumn{1}{c}{\textbf{\makecell{Hyperparameter}}} &
\multicolumn{1}{c}{\textbf{\makecell{Range}}} &
\multicolumn{1}{c}{\textbf{\makecell{Optimized\\Value}}} \\ \hline
\endfirsthead

\hline
\multicolumn{4}{c}{{\tablename} \thetable{} -- Continued} \\ \hline
\multicolumn{1}{c}{\textbf{\makecell{Modelo}}} &
\multicolumn{1}{c}{\textbf{\makecell{Hyperparameter}}} &
\multicolumn{1}{c}{\textbf{\makecell{Range}}} &
\multicolumn{1}{c}{\textbf{\makecell{Optimized\\Value}}} \\ \hline
\endhead

\hline 
\multicolumn{4}{r}{{Continued on next page}} \\ \hline
\endfoot

\hline \hline
\multicolumn{4}{p{0.92\textwidth}}{\small Source: Own elaboration.} \\
\endlastfoot

\textbf{PCA} & Number of Components & 2 a 12 & 6 \\ \hline
\multirow{2}{*}{\textbf{DFM}} 
  & Number of Factors & 2 a 10 & 4 \\
  & Series Length     & 0.01 to 2.5 & 1.2 \\ \hline
\multirow{2}{*}{\textbf{RNA}} 
  & Number of Hidden Layers & 2 to 64  & 32 \\
  & Number of Neurons       & 6 to 256 & 64 \\ \hline
\multirow{3}{*}{\textbf{RNN}} 
  & Number of Hidden Layers & 2 to 48  & 24 \\
  & Number of Neurons       & 6 to 256 & 32 \\
  & Batch Size              & 2.5 to 6.5 & 2.1 \\
\end{longtable}

\begin{figure}[!h]
\centering
\caption{{Private Consumption vs. Optimized Neural Network Indicators}}
\noindent 
\begin{minipage}[c]{.5\textwidth}
  \centering
  \includegraphics[width=1\linewidth]{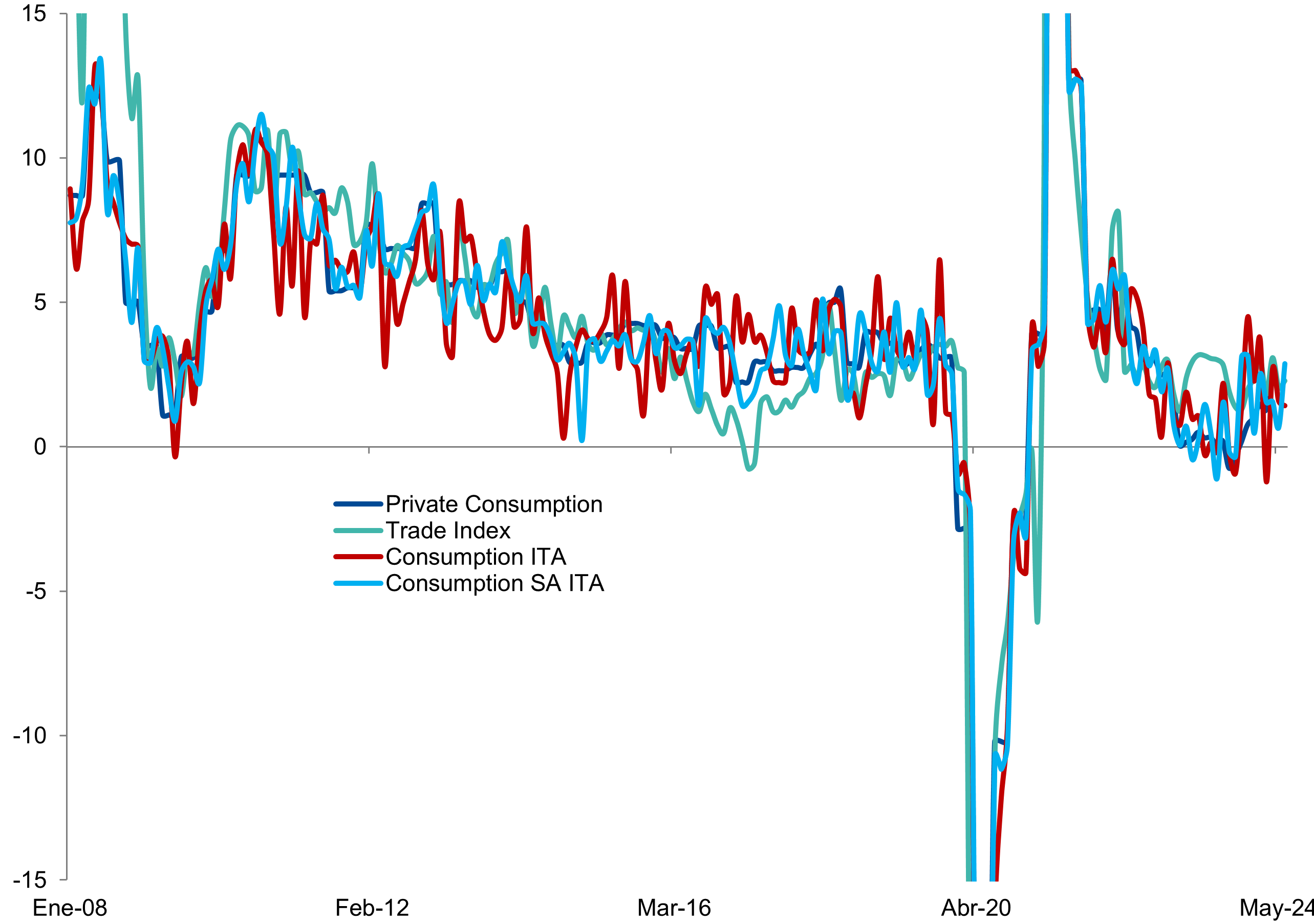}
  \subcaption{Monthly ITAC with Optimal Hyperparameters}\label{fig:1a}
\end{minipage}%
\begin{minipage}[c]{.5\textwidth}
  \centering
  \includegraphics[width=1\linewidth]{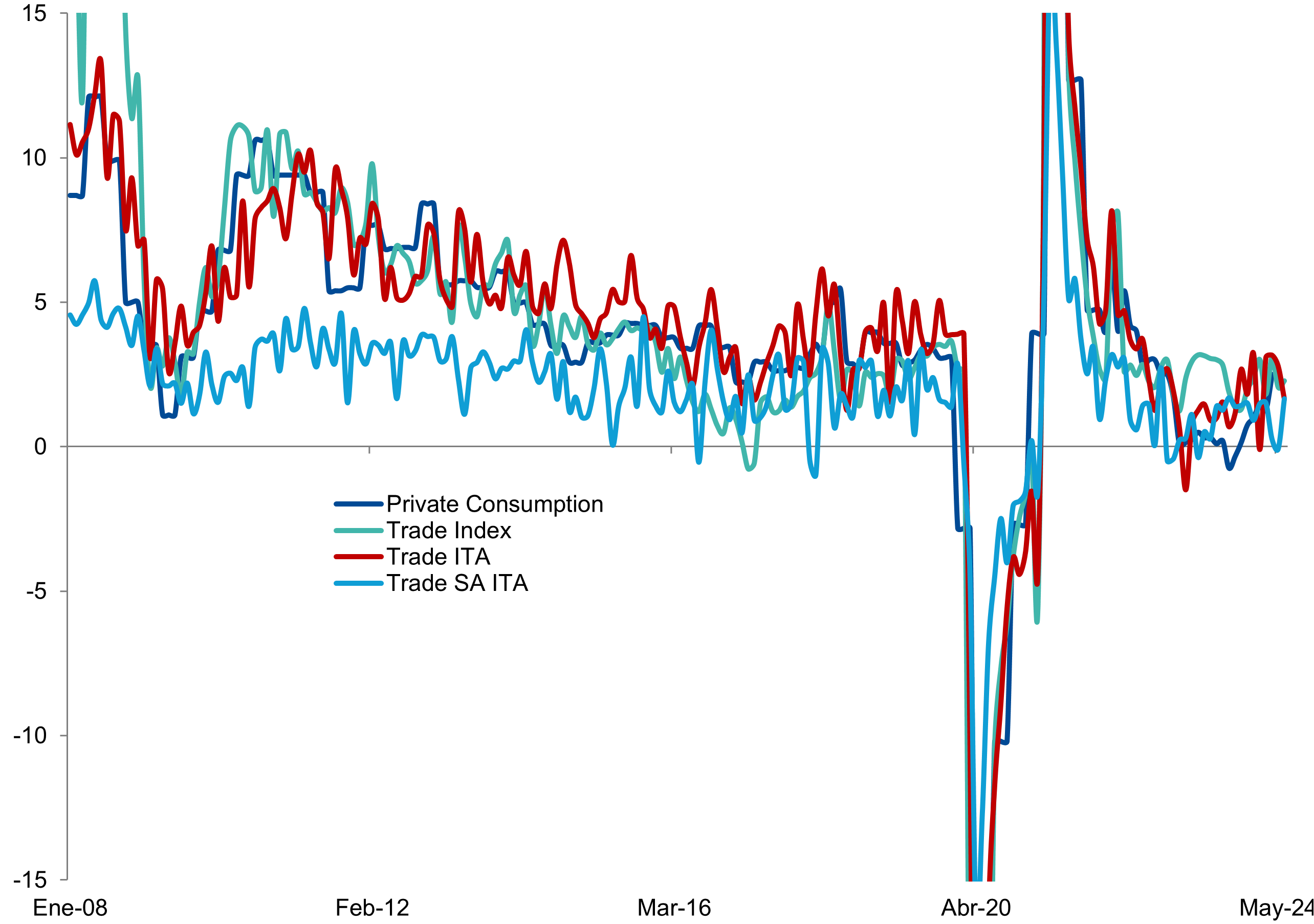}
  \subcaption{Monthly ITAC with Priors}\label{fig:2a}
\end{minipage}

\begin{minipage}[c]{.5\textwidth}
  \centering
  \includegraphics[width=1\linewidth]{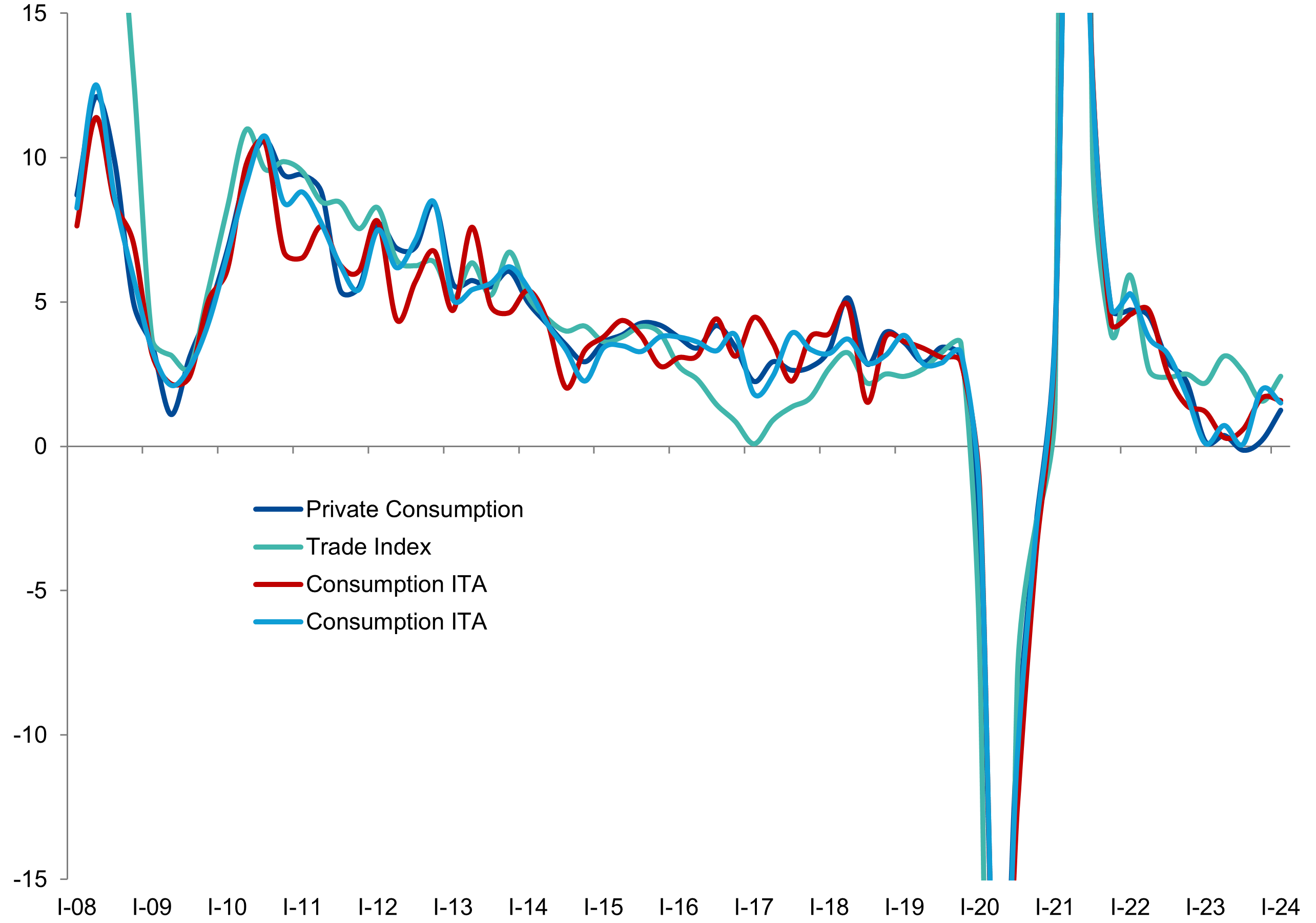}
  \subcaption{Quarterly ITAC with Optimal Hyperparameters}\label{fig:3a}
\end{minipage}%
\begin{minipage}[c]{.5\textwidth}
  \centering
  \includegraphics[width=1\linewidth]{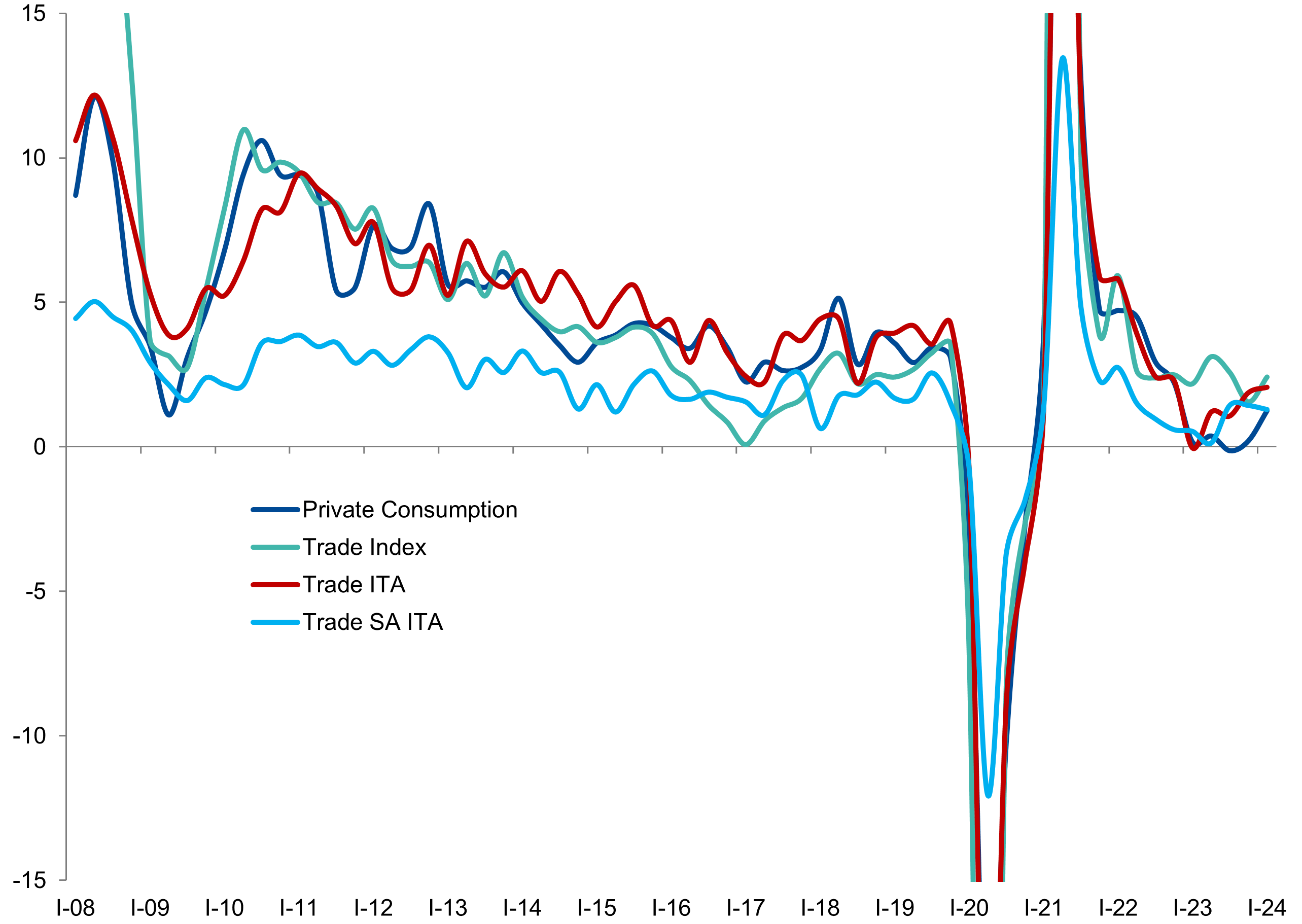}
  \subcaption{Quarterly ITAC with Priors}\label{fig:4a}
\end{minipage}
\raggedright
\footnotesize{Source: Own elaboration}
\end{figure}

\subsection{Robustness Analysis}

In this section, once the best set of models for constructing the ITAC of private consumption and/or trade and services has been obtained, the consistency of the indicator will be evaluated through its predictive capacity (out-of-sample). The forecasting strategy is developed in two stages: (i) an equation is created that includes a set of high-frequency leading indicators theoretically influencing private consumption; the best indicators are selected using the Stepwise Least Squares method; (ii) a state-space equation is formulated where private consumption is the determinant variable, and an ITAC by Google Trends search category is incorporated as a dependent variable based on a selection ranking derived from a Bayesian Variable Selection (BSTS) approach by \cite{scott2013bayesian}. Subsequently, the $X\beta$ component of the first equation is predicted and then included in the second equation. In this way, the second equation seeks to generate predictions by iterating through each search category.

\begin{equation}
    CommeryServics_t = \beta_1 Emplo_t + \beta_2 CreditCons_{t} +\beta_3 CreditHip + \beta_4 CPI + {\epsilon}_t
\end{equation}

\begin{equation}
    Consumption_t = \beta_1 {X}_t + \beta_2 VarGoogle_{it} + e_t
\end{equation}

When analyzing the estimation errors generated by Equation 7, we apply the test proposed by \cite{harvey1997testing}, using a long-run autocorrelation variance estimator based on the method of \cite{diebold1995paring} and  \cite{giacomini2006tests}. This procedure allows for the assessment of improvements in estimation accuracy when correcting the $\beta_1{X}_t$ component.

Table 4 presents the p-values, where the alternative hypothesis states that the models in Equation 8, which incorporate ITAC projections by category, are more accurate than the predictions generated solely by Equation 7. The results indicate a significant improvement in the accuracy of each model that integrates sentiment data, surpassing the 95\% and 99\% confidence levels in a differentiated manner. Notably, the Tourism and Food categories emerge as the best search data sets for predicting private consumption.


\begin{longtable}{l c c c c c}
\caption{Evaluation of Prediction Statistics}
\label{tab:prediction-stats}\\
\hline
\multicolumn{1}{c}{\textbf{Models}} & \multicolumn{1}{c}{\textbf{Estimate}} & \multicolumn{1}{c}{\textbf{MSE}} & \multicolumn{1}{c}{\textbf{RMSE}} & \multicolumn{1}{c}{\textbf{$p$-value (DM)}} & \multicolumn{1}{c}{\textbf{$p$-value (GW)}} \\ \hline
\endfirsthead

\multicolumn{6}{c}{{\tablename} \thetable{} -- Continued} \\
\hline
\multicolumn{1}{c}{\textbf{Models}} & \multicolumn{1}{c}{\textbf{Estimate}} & \multicolumn{1}{c}{\textbf{MSE}} & \multicolumn{1}{c}{\textbf{RMSE}} & \multicolumn{1}{c}{\textbf{$p$-value (DM)}} & \multicolumn{1}{c}{\textbf{$p$-value (GW)}} \\ \hline
\endhead

\hline
\multicolumn{6}{r}{{Continued on next page}} \\
\endfoot

\hline \hline
\multicolumn{6}{p{12cm}}{\small Source: Own elaboration.} \\
\endlastfoot

Food            & 0.614 & 2.35 & 1.40 & 0.029 & 0.002 \\
Transport       & 0.241 & 3.61 & 6.45 & 0.057 & 0.017 \\
Tourism         & 0.795 & 2.02 & 0.98 & 0.005 & 0.032 \\
Recreation      & 0.713 & 4.51 & 5.62 & 0.026 & 0.014 \\
Personal care   & 0.727 & 3.56 & 6.61 & 0.049 & 0.059 \\
Total           & 0.791 & 1.69 & 1.12 & 0.001 & 0.002 \\
\end{longtable}

\section{Conclusions}
\label{sec:conc}

This study highlights the potential of utilizing unstructured data, specifically Google search terms, as a key tool for constructing high-frequency indicators. One of the main advantages identified is the superior ability to detect underlying patterns when sentiment variables are combined with recurrent neural network (RNN) models. This combination allows for the capture of complex and nonlinear dynamics in private consumption behavior, which is particularly relevant in contexts where information is sparse or difficult to obtain. Furthermore, the models' capability to make predictions even with incomplete information represents a significant strength, especially in an environment where much of the available data is fragmented or not fully structured. This aspect is crucial for improving the accuracy of projections in scenarios of uncertainty, where data scarcity is a commonly faced barrier by traditional predictive models. 

Additionally, it is important to emphasize that the incorporation of Google Trends data into neural networks should not be a homogeneous process; it is essential to consider how economic agents interact with search terms, as this may influence the quality of the information provided by these data. In particular, the results of this study demonstrate that a differentiated treatment of Google Trends data for each component of consumption—i.e., adapting the approach according to specific categories—significantly enhances predictive capability. Moreover, the optimal use of Google Trends data varies depending on the forecasting objective, whether to anticipate future consumption or to obtain real-time estimates. This suggests that a uniform approach to integrating Google Trends across different time series and economic contexts does not necessarily yield the best forecasts. Therefore, a specific and tailored adaptation of the treatment of these data is crucial to maximize their value compared to any methodology.

Finally, the implications of this study are highly relevant for both economics and policy formulation. Economists and policymakers have the opportunity to utilize Google Trends data, both daily and weekly, to develop high-frequency indicators that allow for the measurement of key economic variables such as private consumption, employment, and tourism, among others. Given that private consumption represents a significant portion of economic activity, atypical changes in consumption forecasts—especially those correlated with fluctuations in consumption-related search terms or news events—could provide early signals of future large-scale economic events. Furthermore, we share with other economists the necessity of exploring new sources of information, such as Google Trends, which have the potential to become fundamental tools for policymakers. These sources facilitate a more agile and efficient adaptation to an increasingly dynamic economic environment, in which unpredictable events present significant challenges to economic stability.

\newpage

\appendix

\section{Appendix}

\begin{figure}[!ht]
\centering
\begin{minipage}[t]{1\linewidth}
\centering
\caption{{GDP 2000-2023: Percentage Share by Type of Expenditure}}
 \includegraphics[width=0.55\linewidth]{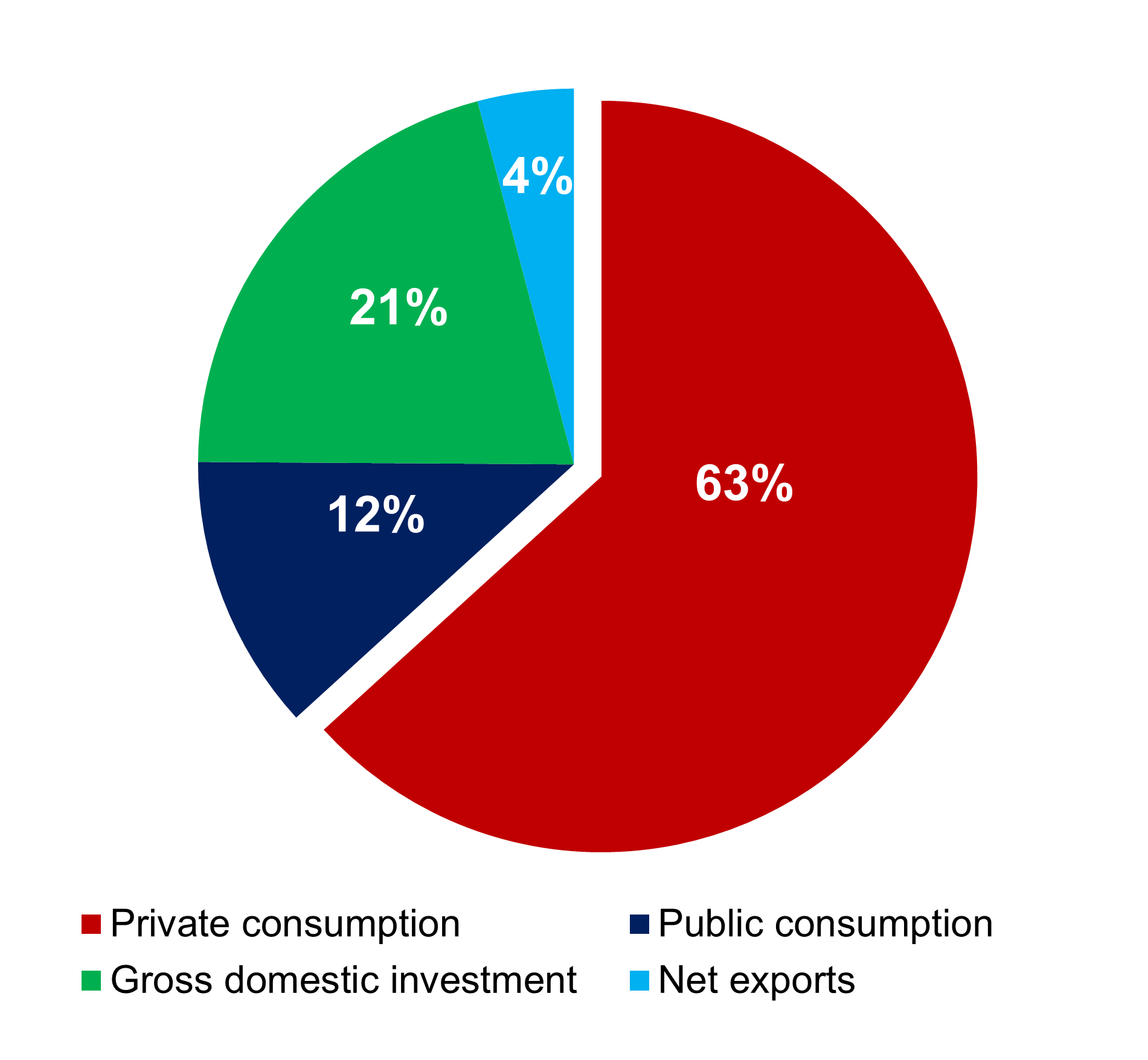}
\end{minipage}\\
\raggedright
\footnotesize{Source: BCRP, own elaboration.}
\end{figure}

\begin{figure}[!ht]
\centering
\begin{minipage}[t]{1\linewidth}
\centering
\caption{{GDP and Contribution Factors 2010-2023}}
 \includegraphics[width=0.7\linewidth]{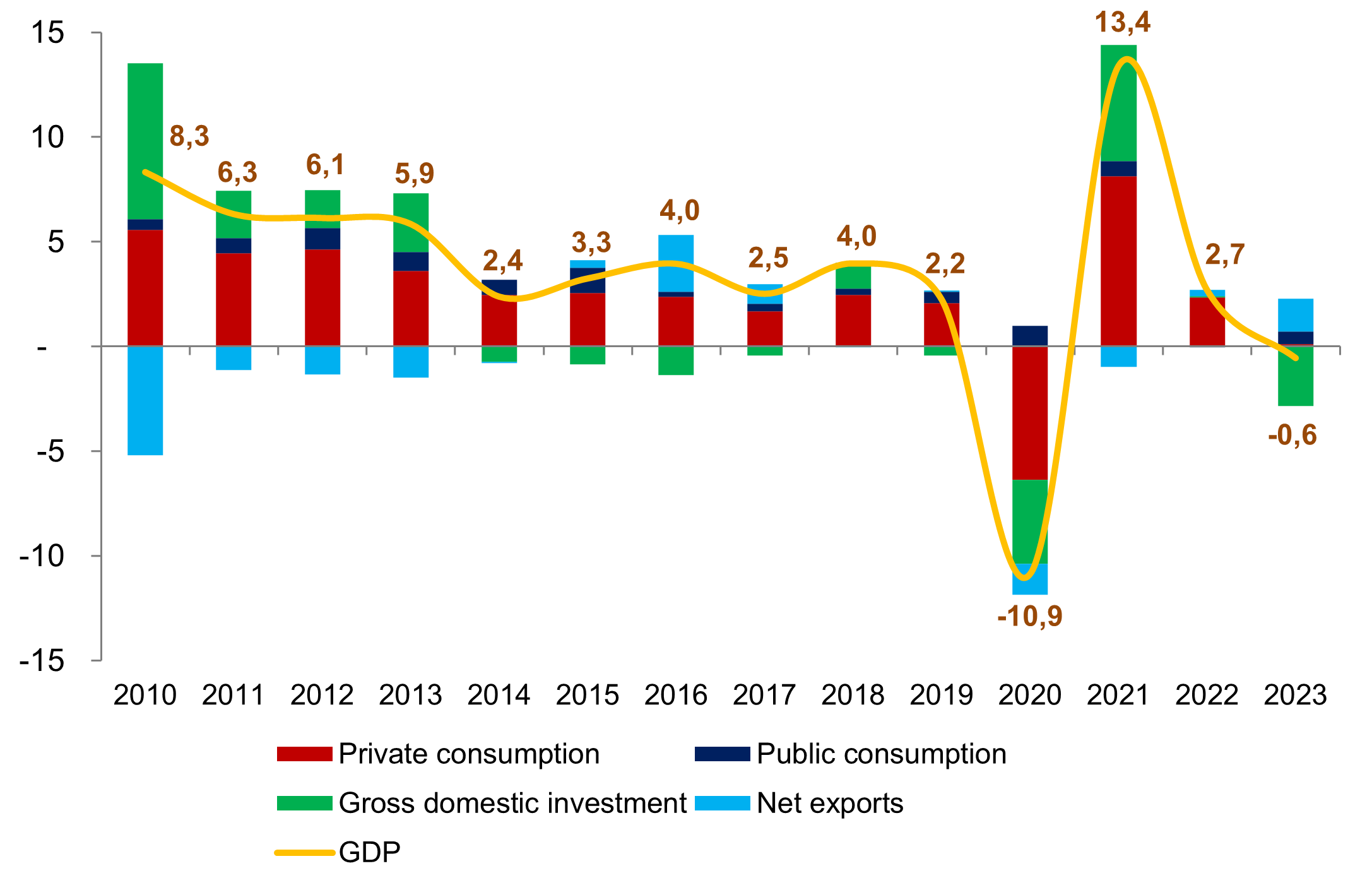}
\end{minipage}\\
\raggedright
\footnotesize{Source: BCRP, own elaboration}
\end{figure}

\newpage


\begin{longtable}{p{2.5cm} p{11.5cm}}
\caption{Total list of tentative search terms}
\label{tab:search-terms}\\
\hline 
\multicolumn{1}{c}{\textbf{Category}} & \multicolumn{1}{c}{\textbf{Words}} \\ \hline
\endfirsthead

\multicolumn{2}{c}{{\tablename} \thetable{} -- Continued} \\
\hline 
\multicolumn{1}{c}{\textbf{Category}} & \multicolumn{1}{c}{\textbf{Words}} \\ \hline
\endhead

\hline 
\multicolumn{2}{r}{{Continued on next page}} \\
\endfoot

\hline \hline
\multicolumn{2}{p{14cm}}{\small Source: Own elaboration.} \\
\endlastfoot

Food & KFC, Pizza Hut, Mc Donald's, Bembos, Wong, Plaza Vea, Metro, Tottus, Rappi, Glovo, Uber Eats, supermarkets, fruits, vegetables, meat, dairy, sodas, beers, wine, commercial centers, stores, restaurants, fast food \\

Tourism & LATAM, SKY, Despegar.com, Cusco, Sheraton, Avianca, Marriot, Hilton, Westin, Machu Picchu, Booking.com, Jorge Chávez, Y tú qué planes, Candelaria Festival, Inti Raymi, Holy Week, Ayacucho, Machu Picchu tickets, Carnival of Cajamarca, tourist tours, tour packages, travel agencies, hotels, domestic flights, bus terminal \\

Services & Movistar hogar, Movistar dúo, Claro hogar, Win internet, electricity bill \\

Home & Sodimac, Maestro, household appliances, electrical appliances, furniture, household products, gardening, furniture stores, decoration, bricolage \\

Personal care & ROE laboratory, Smart Fit, Clínica Internacional, Inkafarma, Mifarma, Soho Color, Montalvo, gym, hospitals, clinic, gimnasio, dance classes, yoga classes, spa, healthy foods, monedero del ahorro \\

Transport & BMV, GLP, GNV, Toyota, Hyundai, Audi, Mazda, Tesla, Indriver, Yango, Cabify, Uber, cars, car insurance, car rentals, taxi service, gas stations \\

Technology & Samsung, Apple, Coolbox, Hiraoka, smartphones, laptops, tablets, televisions, headphones, cellphones \\

Recreation & PSP, Cinemark, Cineplanet, Crisol, Netflix, Disney+, HBO+, Spotify, Cine Star, Black Days, Cyber Days, Comparabien, nightclubs, cinemas, concert, concert tickets, match tickets, video games, books, bookstores, SBS bookstores, theater \\

Education & preschools, universities, schools \\

Finance & AFP withdrawal, Plin, Yape, credit card, loans, bonus, online shopping \\
\end{longtable}


\begin{longtable}{p{3cm} p{5cm} c c}
\caption{List of ITAC search terms}
\label{tab:itac-search-terms} \\
\hline 
\textbf{Category} & \textbf{Words} & \textbf{ITACons} & \textbf{ITACome} \\ \hline
\endfirsthead

\multicolumn{4}{c}{{\tablename} \thetable{} -- Continued} \\
\hline 
\textbf{Category} & \textbf{Words} & \textbf{ITACons} & \textbf{ITACome} \\ \hline
\endhead

\hline 
\multicolumn{4}{r}{{Continued on next page}} \\
\endfoot

\hline \hline
\multicolumn{4}{p{14cm}}{\small Source: Own elaboration.} \\
\endlastfoot

\multirow{2}{*}{Food} 
& restaurants & \cmark & \cmark \\
& Pizza Hut & \xmark & \cmark \\
\hline
\multirow{8}{*}{Transport} 
& car rentals & \cmark & \cmark \\
& Indriver & \cmark & \cmark \\
& GLP & \cmark & \xmark \\ 
& cars & \xmark & \cmark \\ 
& Toyota & \xmark & \cmark \\ 
& Hyundai & \xmark & \cmark \\ 
& Mazda & \xmark & \cmark \\ 
& bus terminal & \cmark & \cmark \\
\hline
\multirow{12}{*}{Tourism} 
& Avianca & \cmark & \cmark \\
& SKY & \cmark & \cmark \\
& hotels & \cmark & \cmark \\
& Marriot & \cmark & \cmark \\
& Jorge Chávez & \cmark & \xmark \\
& tour packages & \cmark & \cmark \\
& Despegar.com & \cmark & \cmark \\
& travel agencies & \cmark & \xmark \\
& Sheraton & \cmark & \xmark \\
& Westin & \cmark & \xmark \\
& Cusco & \xmark & \cmark \\
\hline
\multirow{5}{*}{Recreation} 
& cines & \cmark & \cmark \\
& Cinemark & \cmark & \cmark \\
& theater & \cmark & \cmark \\
& nightclubs & \cmark & \cmark \\
& Cineplanet & \cmark & \xmark \\
\hline
\multirow{6}{*}{Personal care} 
& Smart Fit & \cmark & \cmark \\
& Soho Color & \cmark & \cmark \\
& Spa & \cmark & \cmark \\
& gym & \cmark & \xmark \\
& Montalvo & \cmark & \xmark \\
& gimnasio & \cmark & \xmark \\

\end{longtable}

\begin{longtable}{>{\centering\arraybackslash}p{4cm} >{\centering\arraybackslash}p{4cm} >{\centering\arraybackslash}p{4cm}}
\caption{Private consumption by purpose and ITACons categories}
\label{tab:consumption-itacons}\\
\hline
\multicolumn{1}{c}{\textbf{\makecell{Private consumption\\by purpose of\\INEI}}} & 
\multicolumn{1}{c}{\textbf{\makecell{ITACons\\categories}}} & 
\multicolumn{1}{c}{\textbf{Words}} \\ \hline
\endfirsthead

\hline
\multicolumn{3}{c}{{\tablename} \thetable{} -- Continued} \\ \hline
\multicolumn{1}{c}{\textbf{\makecell{Private consumption\\by purpose of\\INEI}}} & 
\multicolumn{1}{c}{\textbf{\makecell{ITACons\\categories}}} & 
\multicolumn{1}{c}{\textbf{Words}} \\ \hline
\endhead

\hline 
\multicolumn{3}{r}{{Continued on next page}} \\ \hline
\endfoot

\hline \hline
\multicolumn{3}{p{12cm}}{\small Source: INEI, own elaboration.} \\
\endlastfoot

\hline
Food and non-alcoholic beverages / Alcoholic beverages, tobacco, and narcotics 
& Food 
& Pizza Hut, restaurants \\ \hline

Restaurants and hotels 
& Tourism 
& Despegar.com, Cusco, Avianca, SKY, Marriot, tour packages, hotels, bus terminal \\ \hline

Health 
& Personal care 
& Smart Fit, Soho Color, spa \\ \hline

Transport 
& Transport 
& Toyota, Hyundai, Mazda, Indriver, cars, car rentals \\ \hline

Recreation and culture 
& Recreation 
& Cinemark, nightclubs, cines, theater \\ \hline

\end{longtable}


\begin{table}[!htp]
\centering
\caption{Correlations with other private consumption and trade indicators}
\label{tab:correlations}
\begin{adjustbox}{max width=\textwidth, center} 
\begin{tabular}{l c c c c c c c c c c}
\hline
 & \textbf{ITACons} & \textbf{ITACome} & \textbf{VAC} & \textbf{VAOT} & \textbf{VAM} & \textbf{IBC} & \textbf{ISAN} & \textbf{EE3} & \textbf{ES3} & \textbf{ED3} \\
\hline
\textbf{ITACome} & \textbf{0,92} & 1 & & & & & & & & \\
\textbf{VAC}     & \textbf{0,80} & \textbf{0,87} & 1 & & & & & & & \\
\textbf{VAOT}    & \textbf{0,92} & \textbf{0,96} & 0,79 & 1 & & & & & & \\
\textbf{VAM}     & \textbf{0,71} & \textbf{0,79} & 0,88 & 0,72 & 1 & & & & & \\
\textbf{IBC}     & \textbf{0,86} & \textbf{0,93} & 0,86 & 0,92 & 0,88 & 1 & & & & \\
\textbf{ISAN}    & \textbf{0,82} & \textbf{0,80} & 0,80 & 0,79 & 0,74 & 0,87 & 1 & & & \\
\textbf{EE3}     & \textbf{0,54} & \textbf{0,59} & 0,77 & 0,50 & 0,69 & 0,75 & 0,70 & 1 & & \\
\textbf{ES3}     & \textbf{0,59} & \textbf{0,63} & 0,77 & 0,56 & 0,72 & 0,82 & 0,76 & 0,98 & 1 & \\
\textbf{ED3}     & \textbf{0,62} & \textbf{0,66} & 0,81 & 0,60 & 0,76 & 0,84 & 0,77 & 0,96 & 0,98 & 1 \\
\hline
\end{tabular}
\end{adjustbox}

\bigskip
\parbox{\textwidth}{\footnotesize
\textbf{Source:} BCRP, INEI, BBVA, own elaboration. \\
\textbf{Labels:} ITACons = Artificial Trend Index - Consumption, 
ITACome = Artificial Trend Index - Trade, 
VAC = Gross Value Added of trade, 
VAOT = Gross Value Added of other services, 
VAM = Gross Value Added of manufacturing, 
IBC = Consumer Big Data index, 
ISAN = Current business situation index, 
EE3 = Expectations about the economy in 3 months ahead, 
ES3 = Expectations about the sector in 3 months ahead, 
ED3 = Expectations about the demand in 3 months ahead.
}
\end{table}


\begin{figure}[!ht]
\centering
\begin{minipage}[t]{1\linewidth}
\centering
\caption{{Dynamic Correlation of Variables for the ITAC of Private Consumption}}
 \includegraphics[width=.75\linewidth]{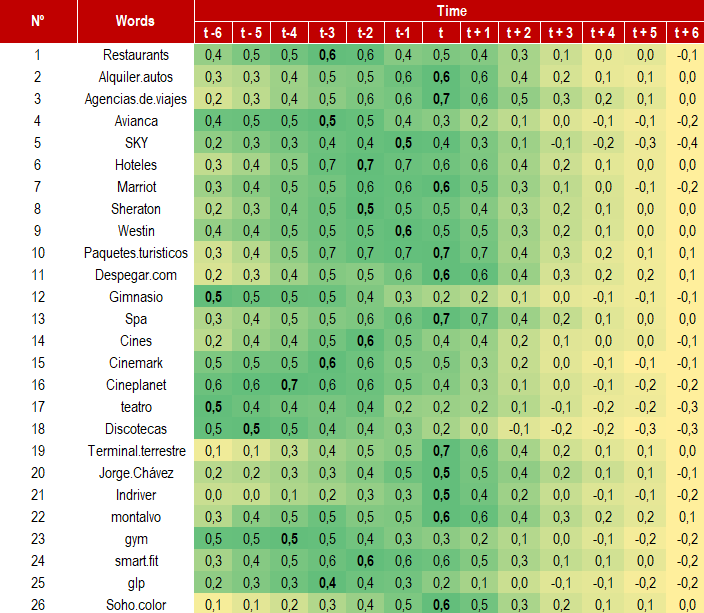}
\end{minipage}\\
\raggedright
\footnotesize{Source: Google Trends, own elaboration}
\end{figure}


\begin{figure}[!ht]
\centering
\begin{minipage}[t]{1\linewidth}
\centering
\caption{{Dynamic Correlation of Variables for the ITAC of Trade and Services}}
 \includegraphics[width=.95\linewidth]{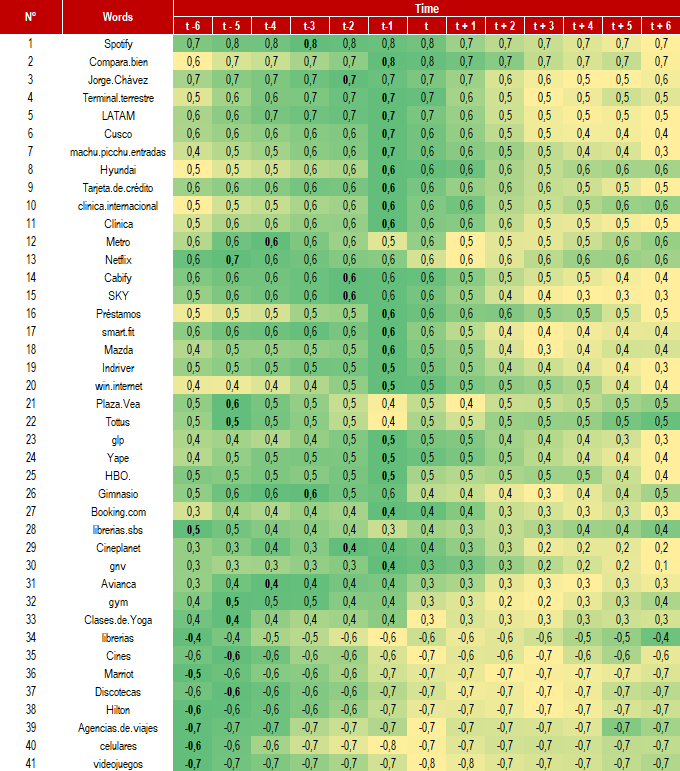}
\end{minipage}\\
\raggedright
\footnotesize{Source: Google Trends, own elaboration}
\end{figure}
\clearpage
\bibliographystyle{elsarticle-num} 
\bibliography{cas-refs}





\end{document}